\documentclass[twocolumn,showpacs,pra,aps]{revtex4}
\usepackage{amsmath}
\usepackage{amsfonts}
\usepackage{amssymb}
\usepackage{graphicx}
\usepackage{natbib}

\newcommand{\be}{\begin{equation}}
\newcommand{\ee}{\end{equation}}

\newcommand{\ket}{\rangle}
\newcommand{\bra}{\langle}

\newlength{\textwidthm}
\begin{document}

\title{Collision induced coherence in plasma and Branching Ratio in multilevel system}

\author{Dong Sun$^{1,2,3}$, Victor V. Kozlov$^{3,4,+}$,Lan Yuan$^{1},$ Yuri V. Rostovtsev$^{3,5}$}

\affiliation {$^{1}$ Department of Optoelectronics Engineering, Xiamen University of Technology, Xiamen 361024, China\\
$^{2}$ Fujian Provincial Key Laboratory of Optoelectronic technologies and devices, Xiamen 361024, China\\
$^{3}$ Department of Physics, Texas A$\&$M University, College Station, TX 77843\\
$^{4}$ Fock Institute of Physics, St. -Petersburg University, Ulyanovskaya 1, St.-Petersburg, Russia, 198904\\
$^{5}$ Department of Physics, University of North Texas, 1155 Union Circle \#311427, Denton, TX 76203 \\
$+$ Deceased}

\date{\today}

\begin{abstract}
Our work is based on the collision-induced coherence of two decay channels along two optical transitions.The quantum interference of pumping
processes creates the dark state and the  more atoms are pumped in this collision-induced dark state the stronger the suppression of the spontaneous
emission. The efficiency of this suppression is quantified by putting it in comparison with  the spontaneous emission on the ultraviolet transition
which proceeds in a regular fashion. The branching ratio of these two(visible and ultraviolet) transitions is introduced as the effective  measure of
the degree of the suppression of the spontaneous emission on the visible transition. Our preliminary calculations show that a significant decrease of
the branching ratio with  increase of electron densities is reproduced in the theory.
\end{abstract}


\maketitle

\section{Introduction}
An excited free atom decays into a vacuum emitting spatially isotropic radiation in the spectrum of frequencies which has Lorentzian shape
with a bandwidth proportional to the Einstein A coefficient\cite{Weisskopf}.The decay rate is a combination of fundamental atomic constants
and strongly resistant to a modification.However,such modifications are possible for instance of the strongly confined atom\cite{Haroche}.
The rate is enhanced or suppressed through the modification of the density of electromagnetic modes in the neighborhood of the resonant
frequency\cite{Purcell}-\cite{Kleppner}. Using a three-level atom driven by coherent fields is another way to control the spontaneous
emission \cite{Cardimona} and observe narrowing of spectral linewidth compared to the natural linewidth\cite{Narducci,Gauthier}. Four-level
driven atomic configurations offer yet another level of control allowing, for example, spectral line elimination and full spontaneous
emission cancelation \cite{SZhu,HLee}.

\begin{figure}[t]
\includegraphics[angle=0,width=0.7\columnwidth]{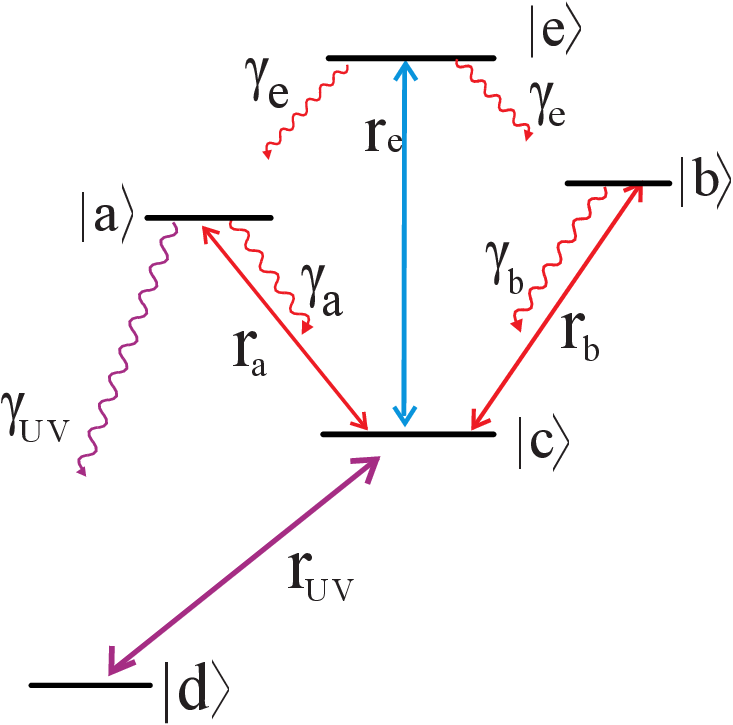}
\caption{Five-level configuration of working levels. Here $\gamma_a$ and $\gamma_b$ are spontaneous decay rates and $r_a$ and $r_b$ are
collision-assisted pumping rates. The $|a\ket\leftrightarrow |d\ket$ is the ultraviolet transition with the spontaneous emission rate
$\gamma_{UV}$. $r_e$ is the pumping rate to state $|e\ket$, $\gamma_e$ is simultaneous equal decay rates to states $|a\ket$ and $|b\ket$.
The rate $r_d$ serves to close the system.} \label{scheme}
\end{figure}

The topic of above-mentioned studies has been focused on individual transitions in atoms and addressed the modification of one or another rate of spontaneous
radiative decay. A number of studies consider more involved schemes characterized by correlated emission simultaneously from two atomic upper levels. Such type
of interference of decay channels has been exploited, for instance, by Harris in his proposal of lasing without inversion in \cite{SHarris}. In contrast to
most studies, we consider not the direct interference of decay channels, but rather the interference of incoherent pumping rates (namely, collision-induced
interference). In its turn, this interference induces the correlation in spontaneous emission. We propose using branching ratio R as the measure of this
correlation. The term branching ratio here stands for the ratio of two spontaneous decay rates from two upper levels $|a\ket$ and $|b\ket$ to two lower
levels $|c\ket$ and $|d\ket$ in a five-level atom of Fig.\ref{scheme}:

\be R=\frac{\gamma_{vis}}{\gamma_{UV}} \label{Ratio1}\ee
where for definition we formulate the branching ratio in terms of decay rates on visible, $\gamma_{vis}$, and ultraviolet,$\gamma_{UV}$, transitions. Here, we
refer to particular transitions in order to keep a close link to the previous experiments on measurements of changes of branching ratios. These experiments
were performed in plasmas in Princeton by Suckewer and coworkers \cite{YChung}. They serve as the main motivation and reference point of our study, although
our results are applicable to more general schemes.

The branching ratio is closely related to total spectral-line intensities (in photons) for the two transitions, $I_{vis}$ and $I_{UV}$. This relation is simply
the consequence of the fact that the intensities are proportional to corresponding decay rates (for schemes considered so far): $I_{vis} \propto \gamma_{vis}$
and $I_{UV} \propto \gamma_{UV}$. In fact, this proportionality allows us to consider the intensity as alternative operational definition of the decay rate
which differs from the pure decay rate by a factor. The access to this factor can be, however, a difficult experimental task. This potential experimental
problem disappears when we consider the branching ratio. In those cases where the radiation is emitted from same level(s) the factors are the same for both
transitions and therefore cancels out in the ratio in Eq. (\ref{Ratio1}). So, the operational definition of the branching ratio can be introduced as \be
R=\frac{I_{vis}}{I_{UV}} \label{Ratio2}\ee

Moreover, for some cases, including ours, the operational definition in Eq. (\ref{Ratio2}) is very natural and probably the only possible definition. The
problem is that we consider simultaneous decay along two (visible) transitions. This decay is characterized by more than one decay rate. This problem does not
appear for the ultraviolet radiation, the decay of which happens along one transition and is therefore characterized by a single decay rate. However, we cannot
formulate the branching ratio in the form of Eq. (\ref{Ratio1}); and instead, we characterize the spontaneous emission process by the operational definition
given by Eq. (\ref{Ratio2}).

For optically thin samples and decays from common upper level, the equality between two definitions Eqs. (\ref{Ratio1}) and (\ref{Ratio2}), i.e. the equality
between the ratio of line intensities and the ratio of the corresponding decay rates, points to the way of measurements of decay rates. Thus, from measurements
of the intensity ratio of two lines, the ratio R and one of the spontaneous decay rates can be deduced knowing the other. Such spectroscopic technique is
rather popular in plasma physics. Particularly attractive is the fact that the branching ratio is as immune to environmental conditions as the decay rate is.
For instance, the ratio does not depend on the electron density and temperature. In the experiments of \cite{YChung}, it is reported that for some transitions
the branching ratio becomes the function of density (concentration). They interpreted this observation as quenching of spontaneous-emission coefficients for
visible transitions. Our study is devoted to formulation of the problem in quantum-optical terms, and our model is based on the five-level scheme (see Fig.
\ref{scheme}), and the effect is caused by collision-induced quantum coherence between two upper levels.

The best way to formulate our main result is to compare two cases. The first one corresponds to no coherence between upper levels a and b. Then, the branching
ratio for visible and ultraviolet transition found from the operational definition (\ref{Ratio2}) reads \be R=2\frac{\gamma_{vis}}{\gamma_{UV}}
\label{Ratio3}\ee as found in the following sections. Here the difference in the factor of two from original definition (\ref{Ratio1}) arises due to doubling
the visible decay rate (here $\gamma_a = \gamma_b \equiv \gamma_{vis}$) by taking into account decays from both upper levels (with equal rates). In contrast,
the rate on the ultraviolet transition is not doubled because the spontaneous decay originates from single level a. Apart from this factor, the equation
(\ref{Ratio3}) stands in line with regular expectations. In particular, this result is unable to explain the concentration dependence of R reported in
\cite{YChung}.

The second case realizes when the (maximal) coherence between levels a and b is induced by electron collisions. Then, as we shall show in the following
sections, the branching ratio under steady-state conditions is given by \be
R=\gamma_{vis}\left(\frac{4}{r_e}+\frac{1}{r_{vis}}\right)=\frac{\gamma_{vis}}{N}\sqrt{\frac{\pi M}{kT}}\left(
\frac{2e^{\frac{E_e}{kT}}}{\overline{k}_e}+\frac{e^{\frac{E_a}{kT}}}{2\overline{k}_{vis}} \right) \label{Ratio4}\ee
where, for simplicity, the result is evaluated in the limit of large pumping rates: $r_{vis}, r_e \gg \gamma_{vis}, \gamma_{UV}$; and $k_i$ is the appropriate
cross-sections of electronic excitation. Since all involved pumping rates depend on the electron concentration, the branching ratio also becomes
concentration-dependent. The higher the concentration the faster the pumping rates and therefore the lower the branching ratio. This relation implies quenching
of spontaneous emission on the visible transition.

In the following sections we show that the effect of quenching is explained by creation of the collision-induced (i.e. induced by pumping
rate $r_{vis}$) dark state as a linear combination of upper states a and b. By formulating the branching ratio R, all common dependencies
cancel out and only the purified asymmetry of responses on the visible and ultraviolet transitions is left. Here, the branching ratio shows
up as a valuable measure of the coherence-induced quenching of spontaneous emission on the visible transition.

\section{Basic set of equations}

We study the set of equations of motion governing temporal evolution of the density matrix elements for the five-level scheme shown in Fig. \ref{scheme}.
The key ingredient of this scheme is the three-level V -type system consisting of two upper states $|a\ket$, $|b\ket$ connected to the lower state $|c\ket$ by
dipole allowed transitions. We first consider this three-level subsystem and then add two states $|d\ket$ and $|e\ket$. We assume that two transitions
$|a\ket-|c\ket$ and $|b\ket-|c\ket$ have close transition frequencies and therefore upper states decay to same continuum of vacuum modes with decay constants
$\gamma_a$ and $\gamma_b$. These constants as well as details of the decay process follow from the Hamiltonian
\be \begin{split} V_{\gamma}&=\hbar \sum_k g_k^{(a)}e^{i(\omega_{ac}-\nu_k)t}|a\ket\bra c|\hat{a}_k\\
&+\hbar \sum_k g_k^{(b)}e^{i(\omega_{bc}-\nu_k)t}|b\ket\bra c|\hat{a}_k+H.c.
\end{split}
\ee describing interaction between the atom and the reservoir of vacuum oscillators, each of frequency $\nu_k$ ($k$ here represents both the momentum
and  polarization of the vacuum mode). Here $g^{(a,b)}_k$ are the coupling constants between the $k$-th vacuum mode and the atomic transitions from
$|a\ket$ and $|b\ket$ to $|c\ket$. $\hat{a}_k$($\hat{a}_k^\dagger$) is the annihilation (creation) operator of a photon in the $k$-th vacuum mode,
which obeys conventional bosonic commutation rule $[\hat{a}_k, \hat{a}_k^\dagger] = \delta_{kk'}$ .

Electron collisions transfer atoms (also ions as in the plasma experiments in Ref. \cite{YChung}) between states in the manner very much like that in
the  process of interaction with an optical field. As a consequence, the Hamiltonian appears in a similar form, as \be V_R=\wp_{ac}\Omega|a\ket\bra
c|+\wp_{bc}\Omega|b\ket\bra c|+H.c.\ee In contrast to coherent optical fields, individual collisional events are not correlated with each other and
therefore the effective field $\Omega$ is incoherent. Note that spectrum of this incoherent field is very broad so that it covers both upper levels
simultaneously, and therefore one and the same field drives both transitions. These transitions are characterized by dipole matrix elements
$\wp_{ac}$ and $\wp_{bc}$ and transition frequencies $\omega_{ac}$ and $\omega_{bc}$. The incoherent field has $\delta$-like correlations at
different times, i.e., \be \bra \Omega^*(t)\Omega(t')\ket\propto\delta(t-t')\ee These correlations are not expected to cover the entire range of
frequencies from $-\infty$ to $+\infty$. It is sufficient that they are at least approximately valid in the vicinity of both resonances and cover the
frequency separation of two upper levels.

The total interaction picture Hamiltonian for the three-level subsystem is the sum of two terms introduced above, \be H=V_\gamma+V_R \ee Derivation
of  equations of motion for the density matrix elements from the total Hamiltonian is a straightforward task. It is based on the scheme well
developed in quantum optics, see for instance \cite{MScully}, where main ingredients are the Wigner-Weisskopf approximation and the generalized
reservoir theory \cite{Fleischhauer}. More details, particularly on the system under consideration, can be found in our paper \cite{Kozlov}. With
this reference to prior works we skip the derivation and jump directly to the equations of motion. They read
\begin{align}
\dot{\rho}_{aa}&=-(r_a+\gamma_a)\rho_{aa}+r_a\rho_{cc}-\frac{1}{2}p\sqrt{r_ar_b}(\rho_{ba}+\rho_{ab})\label{raa}\\
\dot{\rho}_{bb}&=-(r_b+\gamma_b)\rho_{bb}+r_b\rho_{cc}-\frac{1}{2}p\sqrt{r_ar_b}(\rho_{ba}+\rho_{ab})\label{rbb}\\
\dot{\rho}_{cc}&=(r_a+\gamma_a)\rho_{aa}+(r_b+\gamma_b)\rho_{bb}-(r_a+r_b)\rho_{cc}\\
               &+p\sqrt{r_ar_b}(\rho_{ba}+\rho_{ab})\label{rcc}\\
\dot{\rho}_{ab}&=-\frac{1}{2}(r_a+r_b+\gamma_a+\gamma_b)\rho_{ab}-i\Delta\rho_{ab}\\
               &+\frac{1}{2}p\sqrt{r_ar_b}(2\rho_{cc}-\rho_{aa}-\rho_{bb})\label{rab}\\
\dot{\rho}_{ca}&=-\frac{1}{2}(2r_a+r_b+\gamma_a)\rho_{ca}-\frac{1}{2}p\sqrt{r_ar_b}\rho_{cb}\label{rca}\\
\dot{\rho}_{cb}&=-\frac{1}{2}(2r_b+r_a+\gamma_b)\rho_{cb}-\frac{1}{2}p\sqrt{r_ar_b}\rho_{ca}\label{rcb}
\end{align}
Here $r_{a,b} \equiv 2(\wp^2_{ac,bc}/\hbar^2)R$ are pumping rates of atoms to upper states $|a\ket$ and $|b\ket$ induced by collisions with
electrons,  see Fig. \ref{scheme}. Detuning $\Delta = \omega_{ac}-\omega_{bc}$ is supposed to be small as compared to the optical frequency and
should be of the order of pumping rates $r_a$ or $r_b$ or less in order to make quantum interference effects observable.

Terms containing products of pumping rates appear due to the interference of two optical transitions which is induced by collisions with electrons.
These terms are central to our discussion. In order to emphasize the role of this collision-induced interference, we assume that decay channels do
not interfere. The interference terms are accompanied by the $p$ factor. This factor is the normalized scalar product of corresponding dipole matrix
elements: \be p=\frac{\wp_{ac}\cdot\wp_{bc}}{|\wp_{ac}\wp_{bc}|}\ee

According to its definition, the alignment factor takes the value equal to $1$ for parallel dipole moments, $-1$ for antiparallel, and $0$ for orthogonal.
Intermediate values on $[-1, 1]$ segment are also possible. Maximal coherence corresponds to parallel or antiparallel dipole moments, while zero coherence
corresponds to orthogonal dipole moments. These two extremes of maximal and minimal coherence deserve special attention.

It is not difficult to proceed with modeling the system shown in Fig. \ref{scheme} and supplement the three-level subsystem with two additional levels. For
further considerations we shall need only equations for populations and one equation for the polarization on $|a\ket-|d\ket$ transition. The upper state
$|e\ket$ is subject to decay with rate $\gamma_e$ and pumping from state $|c\ket$ by electronic collisions with rate $r_e$. So, for the population we
immediately obtain \be \dot{\rho}_{ee}=r_e(\rho_{cc}-\rho_{ee})-2\gamma_e\rho_{ee} \label{ree}\ee

Of course, we could derive this equation from first principles by writing the Hamiltonian of interaction of the atom with the continuum of quantum oscillators
to model the decay and interaction with the incoherent electron field to model the pumping process. Since no interference effects are expected on this
transition, this derivation is no more than a trivial exercise. The intuitive appearance of this equation stays in contrast to the detailed consideration of
the V -type subsystem where the nontrivial appearance of interference terms requires cautious analysis.

Similar equation can be written for the lowest state of the system, $|d\ket$. Here the decay with rate $\gamma_{UV}$ from $|a\ket$ state pumps the $|d\ket$
state. The electron pumping is modeled by rate $r_{UV}$. So, in accordance with the picture in Fig. \ref{scheme} we write \be
\dot{\rho}_{dd}=r_{UV}(\rho_{cc}-\rho_{dd})+\gamma_{UV}\rho_{aa}\ee

We now complete the set of equations by writing the equation of motion for the polarization on $|a\ket-|d\ket$ transition: \be
\dot{\rho}_{ad}=-\frac{1}{2}(\gamma_{UV}+r_{UV})\rho_{ad}\ee

We can also write down equations of motion for the other polarizations involving states $|e\ket-|d\ket$. However, we do not do this because these equations are
not used in further analysis. For more general considerations, if necessary, they can be composed by following directly the picture in Fig.\ref{scheme}.
Note that no interference terms appear.

To complete the model we should account for decays and pumpings between the $V$-subsystem and $|d\ket$ and $|e\ket$ states. The subset of equations
(\ref{raa})-(\ref{rcb}) are modified in the following way. There is one incoming term for the population of $|b\ket$ state: $+\gamma_e\rho_{ee}$. One incoming
and one outcoming term for $\rho_{aa}: +\gamma_e\rho_{ee}$ and $-\gamma_{UV}\rho_{aa}$. For $\rho_{cc}$ we get two additional pumping terms,
$+r_e(\rho_{ee}-\rho_{cc})$ and $+r_{UV}(\rho_{dd}-\rho_{cc})$. The equation of motion for the two-photon coherence $\rho_{ab}$ should be supplemented with
decay term $\frac{1}{2}\gamma_{UV}\rho_{ab}$. One-photon polarization $\rho_{ac}$ also decays faster due to additional decay term $\frac{1}{2}(r_e + r_{UV} +
\gamma_{UV})\rho_{ac}$. Finally, in the equation for $\rho_{bc}$ we should also write additional decay term $\frac{1}{2}(r_e + r_{UV})\rho_{bc}$. In the next
section we summarize all these changes and after implementing a few simplified assumptions write down modified equations.

\section{Dressed-state analysis}
We consider properties of the V-type subsystem $|a\ket-|b\ket-|c\ket$ in the dressed-state picture. We assume equal dipole moments on
$|a\ket-|c\ket$ and $|b\ket-|c\ket$ transitions for simplifications: $\wp_{ac} = \wp_{bc}$. This assumption immediately yields the equality
of decay constants: $\gamma_a = \gamma_b$. We also get equal pumping rates, $r_a = r_b$. Also, let the two transitions have equal
frequencies, so that two-photon detuning $\Delta$ is zero.

The upper level $|e\ket$ is auxilary with respect to the rest of the system. It serves to model the external pumping of two upper levels of the V -subsystem.
In plasmas this pumping is due to thermal redistribution of population from all levels which are higher with respect to $|a\ket$ and $|b\ket$. In order to
accurately account for details of this process we would have to consider a particular level scheme for a given atom/ion. However, our immediate goal is not the
one-to-one description of a particular experiment but rather the proof-of-principle theoretical study of relevant coherent effects. Guided by simplicity we
eliminate level $|e\ket$ from our five-level scheme. For that purpose, we assume fast decay rate $\gamma_e$ with respect to pumping rate $r_e$. Then, from the
equation (\ref{ree}) we can write for the population $\rho_{ee}$ in steady-state: \be \rho_{ee}\approx \frac{r_e}{\gamma_e}\rho_{cc}\ee

With these simplifications in effect and additional terms from the discussion in the end of the previous section, equations (\ref{raa})-(\ref{rcb}) for the V
-subsystem become
\begin{align}
\dot{\rho}_{aa}&=-(r_{vis}+\gamma_{vis}+\gamma_{UV})\rho_{aa}+(r_{vis}+\frac{r_e}{2})\rho_{cc}-p r_{vis}\rho_{ab} \label{nraa} \\
\dot{\rho}_{bb}&=-(r_{vis}+\gamma_{vis})\rho_{bb}+(r_{vis}+\frac{r_e}{2})\rho_{cc}-p r_{vis}\rho_{ab} \label{nrbb}
\end{align}
\be
\begin{aligned}\dot{\rho}_{cc}=&(r_{vis}+\gamma_{vis})(\rho_{aa}+\rho_{bb})-(2r_{vis}+r_e+r_{UV})\rho_{cc}\\
               &+2p r_{vis}\rho_{ab}+r_{UV}\rho_{dd} \label{nrcc}\end{aligned}\ee
\be
\dot{\rho}_{ab}=-(r_{vis}+\gamma_{vis}+\frac{\gamma_{UV}}{2})\rho_{ab}+\frac{1}{2}pr_{vis}(2\rho_{cc}-\rho_{aa}-\rho_{bb})\label{nrab}
\ee
with new notations: $\gamma_a = \gamma_b \equiv \gamma_{vis}$ and $r_a = r_b \equiv r_{vis}$. Here we write down only equations relevant for the study in this
section. Equations for one-photon polarizations $\rho_{ca}$ and $\rho_{cb}$ will be considered in the next section.

Coherent effects in the V-subsystem appear due to the interference of pumping channels along $|a\ket-|c\ket$ and $|b\ket-|c\ket$
transitions. Actually, it is the same (incoherent) pumping process that drives both transitions. Here, the pumping field dresses atomic
states $|a\ket$ and $|b\ket$ yielding the collision-induced dark state and its orthogonal partner-bright state with equal dipole moments
$r_a = r_b$.
\begin{align}
|D\ket &=\frac{1}{\sqrt{2}}(|a\ket-|b\ket) \label{DD}\\
|B\ket &=\frac{1}{\sqrt{2}}(|a\ket+|b\ket) \label{BB}
\end{align}

The equations of motion in terms of the dressed variables requires introduction of corresponding density matrix elements. They are
\begin{align}
\rho_{DD}&=\frac{1}{2}(\rho_{aa}+\rho_{bb}-2\rho_{ab}) \label{rDD}\\
\rho_{BB}&=\frac{1}{2}(\rho_{aa}+\rho_{bb}+2\rho_{ab})\label{rBB}\\
\rho_{DB}&=\frac{1}{2}(\rho_{aa}-\rho_{bb}) \label{rDB}
\end{align}

Given equations of motion in bare-states basis (\ref{nraa})-(\ref{nrab}), and dressed-states definitions (\ref{rDD})-(\ref{rDB}) for the
density matrix elements, we reformulate these equations as \be
\dot{\rho}_{DD}=-\gamma_{vis}\rho_{DD}-\frac{\gamma_{UV}}{2}(\rho_{DD}+\rho_{DB})+r_{e}\rho_{cc} \label{DDD}\ee \be
\dot{\rho}_{BB}=-(2r_{vis}+\gamma_{vis})\rho_{BB}-\frac{\gamma_{UV}}{2}(\rho_{BB}+\rho_{DB})+(2r_{vis}+r_{e})\rho_{cc} \label{DBB} \ee \be
\dot{\rho}_{DB}=-(r_{vis}+\gamma_{vis}+\frac{\gamma_{UV}}{2})\rho_{DB}-\frac{1}{4}\gamma_{UV}(\rho_{DD}+\rho_{BB})\label{DDB} \ee

Our ultimate goal is the comparison of intensities of spontaneous emissions on the visible transition and on the ultraviolet transition. As
we shall derive in the next section, these intensities are linearly proportional to the population of $|B\ket$ state and $|a\ket$ state,
correspondingly. Both these populations can be obtained from the solution of equations (\ref{DDD})-(\ref{DDB}) and inverted conversion
formulas (\ref{rDD})-(\ref{rDB}). It is not difficult to obtain exact steady-state solutions for all three dressed-states density matrix
elements in above equations. However, we prefer to make yet another simplification in order to bring physics in the clearest possible form.

We assume that pumping rates are much faster than decay processes, particularly faster than the fastest decay rate $\gamma_{UV}$. Note that the spontaneous
emission rate $\gamma_{UV}$ is two orders of magnitude greater than decay rate $\gamma_{vis}$ on the visible transition. Clearly, this difference is attributed
to the cubic dependence of spontaneous decay rates on the transition frequency. In its turn, the high efficiency of pumping processes arises from frequent
collisions of ions with free electrons (note that the electron concentration $N_e$ exceeds the value of $10^{18} cm^{-3}$). Finally, we formulate our
assumption in the form of strong inequality \be r_{vis},r_{e} \gg \gamma_{UV},\gamma_{vis} \label{rrgg}\ee

This inequality is the key to understanding the effect of suppression of spontaneous emission on the visible transition. This understanding comes from the
analysis of the equation (\ref{DDD}). We shall see that more atoms in the dark state makes the emission weaker. The first two terms in equation (\ref{DDD})
stand for the depletion of the dark state and are therefore undesirable for our purposes. [Note that steady-state polarization $\rho_{DB}$ contributes as
little as $(\gamma_{UV}/r_{vis})\rho_{DD}$.] This depletion is counterbalanced by the gain associated with the last term. The large value of this term requires
fast pumping rate that is guaranteed by inequality (\ref{rrgg}).

With inequality (\ref{rrgg}) we get approximate solutions of equations (\ref{DDD})-(\ref{DDB}). They are
\begin{align}
\rho_{DD} &\approx \frac{r_e}{\gamma_{UV}}\rho_{cc} \label{DDc} \\
\rho_{BB} &\approx (1+\frac{r_e}{4r_{vis}})\rho_{cc} \label{BBc} \\
\rho_{DB} &\approx \frac{r_e}{r_{vis}}\rho_{cc} \label{DBc}
\end{align}
where we
also used $\gamma_{UV} \gg \gamma_{vis}$. All these solutions are expressed in terms of $\rho_{cc}$ which remains unknown quantity until we solve all equations
of motion for the closed five-level system. However, the explicit knowledge of $\rho_{cc}$ is not necessary for our study.

Solution (\ref{DDc}) shows that the dark state is populated only due to the pumping via auxiliary upper state $|e\ket$. By comparing solutions given by
formulas (\ref{DDc}) and (\ref{BBc}) and using inequality (\ref{rrgg}), we conclude that $\rho_{DD} \gg \rho_{BB}$, so that the population of upper states
$|a\ket$ and $|b\ket$ is dominantly concentrated in the dark state. This asymmetry means that most atoms are trapped in the non-emitting state $|D\ket$. As we
shall see shortly, the spontaneous emission on the visible transition is proportional to $\rho_{BB}$; and therefore basing on the just derived strong
inequality $\rho_{DD} \gg \gamma_{BB}$, we conclude that the emission is strongly suppressed. This suppression is to be put in comparison with the regular
(unsuppressed) emission on the ultraviolet transition, which is regulated by the amount of population in the $|a\ket$ state. Overall, the degree of suppression
is quantitatively estimated by the corresponding branching ratio, as the ratio of two emissions, see Eq. (\ref{Ratio2}) and derivations in the next section.

The population of the $|a\ket$ state as well as $\rho_{bb}$ and $\rho_{ab}$ can be found by inverting definitions (\ref{rDD})-(\ref{rDB}). So, in terms of bare
states we get
\begin{align}
\rho_{aa} &=\frac{1}{2}(\rho_{BB}+\rho_{DD}+2\rho_{DB})\approx \frac{r_e}{2\gamma_{UV}}\rho_{cc} \label{rraa} \\
\rho_{bb} &=\frac{1}{2}(\rho_{BB}+\rho_{DD}-2\rho_{DB})\approx \frac{r_e}{2\gamma_{UV}}\rho_{cc} \label{rrbb} \\
\rho_{ab} &=\frac{1}{2}(\rho_{BB}-\rho_{DD})\approx -\frac{r_e}{2\gamma_{UV}}\rho_{cc} \label{rrab}
\end{align}
This state is the state of maximal coherence in the sense that $|\rho_{ab}| \approx \sqrt{\rho_{aa}\rho_{bb}}$. This coherence is of
collisional nature and its high degree is the reflection of the high efficiency of the pumping mechanism ($r_{vis}$).

\section{Spontaneous emission spectra}

We calculate the spectrum $S_{vis}(\omega)$ of spontaneously emitted photons on the visible transition, i.e. through $|a\ket-|c\ket$ and
$|b\ket-|c\ket$ channels. Then, we similarly calculate the ultraviolet spectrum $S_{UV}(\omega)$ of photons emitted in the $|a\ket-|d\ket$
channel. Integrating these two spectra over the frequencies in the vicinity of the corresponding emission peaks and normalizing to the
energy of one photon, we thus obtain two spectral-line intensities \be I_{vis}=(\hbar\omega_{vis})^{-1}\int d\omega S_{vis}(\omega)\ee \be
I_{UV}=(\hbar\omega_{UV})^{-1}\int d\omega S_{vis}(\omega)\ee

In the case of three levels the calculation procedure quickly becomes cumbersome and requires numerical analysis on the final
stage\cite{LMnarducci}. In our case the system is driven by an incoherent field that makes the situation a little simpler. Furthermore,
applying the so far discussed approximations and simplifications, we can make the problem even analytically tractable with transparent
final result. We even bypass the direct application of the quantum regression theorem.

First, we take the definition of the spectrum (valid for stationary processes) as the Fourier transform of the autocorrelation function of the second order of
the scattered electric field: \be S(\omega)=\frac{1}{\pi}Re \int^{\infty}_{0}d\tau e^{-i\omega\tau} \bra E^{-}(r,t)\cdot E^{+}(r,t+\tau)\ket \label{S}\ee

This autocorrelation function is simply related to the normally-ordered product of polarizations taken at instants of time separated by a positive time delay
$\tau$. This relationship follows from the well-known formula \be E^{+}(r,t)=-\frac{\omega_0^2}{4\pi\varepsilon_0  c^2 r}
\mathbf{n}\times(\mathbf{n}\times\mathbf{d})\mathbf{P}^{(+)}(t-r/c) \ee connecting the electric field $\mathbf{E}^{(+)}$ with polarization $\mathbf{P}^{(+)}$.
Here, $\mathbf{n}$ is the unit vector in the direction of observation, $\mathbf{d}$ is the unit vector along the atomic dipole moment, $\mathbf{r}$ is the
observation point, measured from the position of the atom, and $\omega_0$ is the polarization frequency. More details can be found, for instance in Ref.
\cite{LMnarducci}. The resultant formulas are
\begin{align} S_{vis}(\omega) &=\frac{C\omega^4_{vis}}{\pi}Re \int^{\infty}_{0}d\tau e^{-i\omega\tau} \bra P^{-}_{vis}(t)\cdot P^{+}_{vis}(t+\tau)\ket \\
S_{UV}(\omega) &=\frac{C\omega^4_{UV}}{\pi}Re \int^{\infty}_{0}d\tau e^{-i\omega\tau} \bra P^{-}_{UV}(t)\cdot P^{+}_{UV}(t+\tau)\ket
\end{align}

Here $C$ is an unimportant constant, same for both transitions. $P^{(-)}_{vis}$ and $P^{(+)}_{vis}$ ($P^{(-)}_{UV}$and $P^{(+)}_{UV}$) are negative and
positive frequency parts of the polarization induced on the visible(ultraviolet) transition. According to our scheme they are defined as \be
P^{-}_{vis}(t)=\wp_{ac}\sigma_{ac}(t)+\wp_{bc}\sigma_{bc}(t)\label {pvn}\ee \be P^{+}_{vis}(t+\tau)=\wp_{ca}\sigma_{ca}(t+\tau)+\wp_{cb}\sigma_{cb}(t+\tau)
\label{pvp}\ee

Similarly we define the negative- and positive-frequency parts of the polarization on the ultraviolet transition: \be P^{-}_{UV}(t)=\wp_{ad}\sigma_{ad}(t)
\label{pun}\ee \be P^{+}_{UV}(t+\tau)=\wp_{da}\sigma_{da}(t+\tau) \label{pup}\ee

In the following we assume $\wp_{ac} = \wp_{ca} = \wp_{bc} = \wp_{cb}$ and $\wp_{ad} = \wp_{da}$.

Operators of atomic transitions $\sigma_{ij}$ are defined in usual way: $\sigma{ij} = |i\ket\bra j|$. They are particularly relevant for the calculation of
multi-time correlation functions. Note that these quantities are quantum-mechanical operators. Their quantum-mechanical average has simple relation to the
elements of the density matrix, namely
\be \begin{aligned} \bra \sigma_{ij}(t)\ket &= Tr[U^+(t)\sigma_{ij}(0)U(t)\rho(0)]\\
                                            &= Tr[\sigma_{ij}(0)U(t)\rho(0)U^+(t)]\\
                                            &= Tr[\sigma_{ij}(0)\rho(t)] \end{aligned} \label{Sij}\ee

Since we need to calculate two-time correlation functions, it is instructive to write down equations of motion for relevant operators and then find solutions
to the initial-value problem. The form of these equations coincide with the equations of motion for respective density matrix elements. This can be checked by
direct derivation from the Heisenberg equation of motion with the Hamiltonian \cite{Gauthier}. Thus we get
\begin{align}
\sigma_{ac} &=-\frac{1}{2}(3r_{vis}+r_{UV}+r_e+\gamma_{vis}+\gamma_{UV})\sigma_{ac}-\frac{1}{2}r_{vis}\sigma_{bc} \label{Sac} \\
\sigma_{bc} &=-\frac{1}{2}(3r_{vis}+r_{UV}+r_e+\gamma_{vis})\sigma_{bc}-\frac{1}{2}r_{vis}\sigma_{ac} \label{Sbc} \\
\sigma_{ad} &=-\frac{1}{2}(r_{vis}+r_{UV}+r_e+\gamma_{vis}+\gamma_{UV})\sigma_{ad} \label{Sad}
\end{align}end{align} where we used the assumption $r_a = r_b \equiv r_{vis}$.
Note that $\sigma_{ji} = \sigma_{ij}^{\dagger}$ and the equations of motion for the conjugate operators arise by taking the Hermite conjugate of the right-hand
sides of Eqs. (\ref{Sac})-(\ref{Sad}). The above equations form the closed system and allow simple solutions. These solutions acquire even simpler form when we
apply our key inequality (\ref{rrgg}). Thus, for the quantities of interest we get \be
\sigma_{ca}(t+\tau)+\sigma_{cb}(t+\tau)=e^{-\frac{1}{2}r_0\tau}[\sigma_{ca}(t)+\sigma_{cb}(t)] \label{Sca}\ee \be \sigma_{da}(t+\tau) =
e^{-\frac{1}{2}(r_{UV}+r_{vis})\tau}\sigma_{da}(t)\label{Sda}\ee where $r_0 \equiv (2r_{vis} + r_{UV} + r_e)$.

Everything is ready for the formulation of the desired result. After substituting solution (\ref{Sca}) in the expression (\ref{pvp}) and correspondingly
(\ref{Sda}) in (\ref{pup}), the two-time correlation functions of the polarizations read
\be
\begin{aligned}\bra P^{-}_{vis}(t)\cdot P^{+}_{vis}(t+\tau)\ket &=\wp^2_{ac}e^{-r_0\tau} \\
&\times \bra \sigma_{aa}(t)+\sigma_{bb}(t)+\sigma_{ab}(t)+\sigma_{ba}(t)\ket \end{aligned}\ee
\be \bra P^{-}_{UV}(t)\cdot P^{+}_{UV}(t+\tau)\ket=\wp^2_{ad}e^{-(r_{UV}+r_{vis})\tau}\bra \sigma_{aa}(t)\ket \ee

The Fourier transform of the above expressions converts the exponential decays in time to the Lorentzian spectra in frequency:
\begin{align}
S_{vis}(\omega) &=\frac{C\omega^2_{vis}\wp_{ac}^2}{\pi}\frac{r_0}{\omega^2+r_0^2}2\rho_{BB}(t)\\
S_{UV}(\omega)  &=\frac{C\omega^2_{UV}\wp_{ad}^2}{\pi}\frac{r_{UV}+r_{vis}}{\omega^2+(r_{UV}+r_{vis})^2}\rho_{aa}(t)
\end{align}
Here we take only the real part of the integral and use the rule (\ref{Sij}) for calculating quantum-mechanical averages, and also the definition of the
population of the bright state (\ref{rBB}). Note also the equality $\rho_{ab} = \rho_{ba}$ valid for the zero value of the two-photon detuning, which is the
case considered here.

Note here that it may seem that the thus obtained spectra can, in principle, become a function of time, copying the time dependence of populations $\rho_{BB}$
and $\rho_{aa}$. This dependence is a signature of a nonstationary process. Therefore the spectrum of these solutions is to be defined not by Eq. (\ref{S}) but
with the more general formula which is derived and discussed for instance in Ref. \cite{MScully}. However, in our case the interest is in stationary emission
corresponding to steady-state solutions for $\rho_{BB}$ and $\rho_{aa}$. Therefore, $\rho_{BB}$ and $\rho_{aa}$ become time independent. In Fourier space they
are simply two stationary Lorentzian peaks. Spectral-line intensities follow as the frequency integral over these peaks:
\begin{align}
I_{vis}&=\hbar^{-1}C\omega^3_{vis}\wp^2_{ac}2\rho_{BB}\\
I_{UV} &=\hbar^{-1}C\omega^3_{UV}\wp^2_{ad}\rho_{aa}
\end{align}

The branching ratio is the ratio of two above quantities, as regulated by definition (\ref{Ratio2}): \be
R=\frac{\gamma_{vis}}{\gamma_{UV}}\frac{2\rho_{BB}}{\rho_{aa}} \label{Ratio5}\ee where we used the definition of the spontaneous decay constant \be
\gamma_{vis}=\frac{1}{4\pi\varepsilon_0}\frac{4\omega^3_{vis}\wp^2_{ac}}{3\hbar c^3} \label{gam}\ee and similar formula for the ultraviolet transition, and
canceled out common prefactors.

This ratio is the main formula of our chapter. It reflects the role of coherent effects in the spontaneous emission along the visible transition. A small
population in the bright state on the background of relatively strongly populated $|a\ket$ state implies the suppression of the spontaneous decay. Such
asymmetry becomes possible only when the coherence $\rho_{ab}$ between the two upper states $|a\ket$ and $|b\ket$ acquires an appreciable value. Here, the
coherence is induced by collisions and its large value is guaranteed by the inequality (34), which means that decay processes destroying the coherence are of
little importance. Quantitatively, the degree of the suppression is related to the value of R in absence of the interference terms which becomes simply the
double ratio of decay constants, as we show below.

\section{Suppression of the spontaneous decay}

\begin{figure}[t]
\includegraphics[width=0.8\columnwidth]{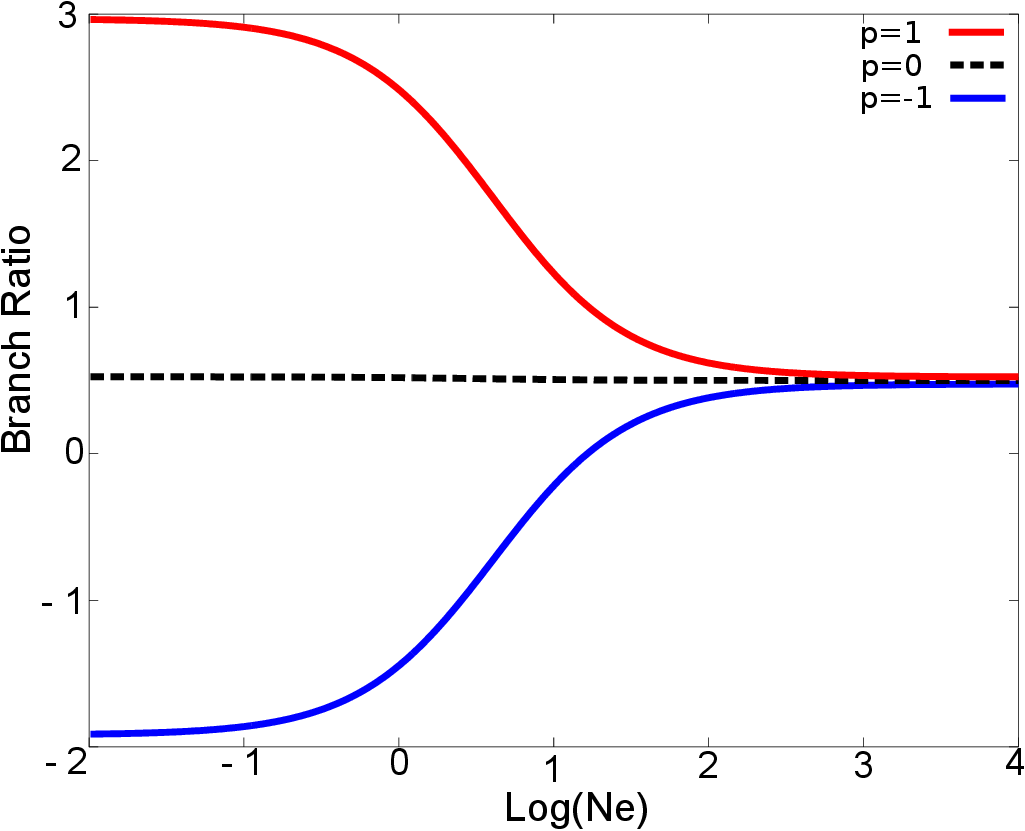}
\includegraphics[width=0.8\columnwidth]{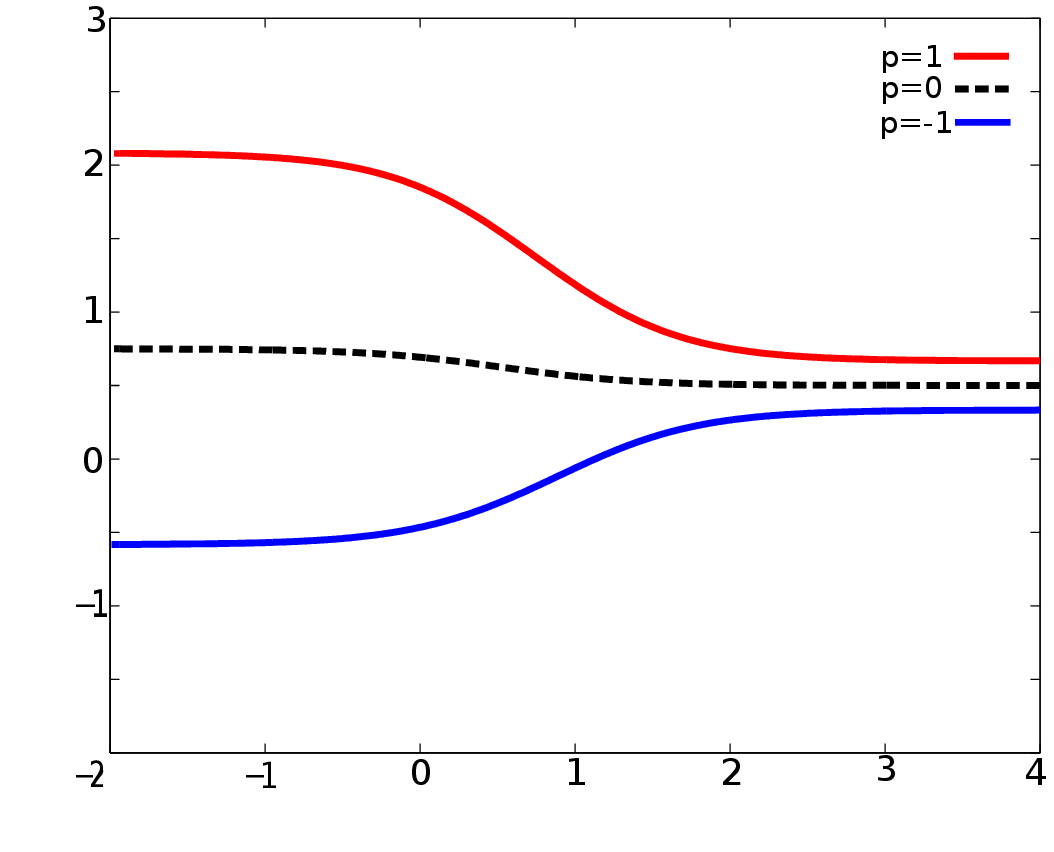}
\includegraphics[width=0.8\columnwidth]{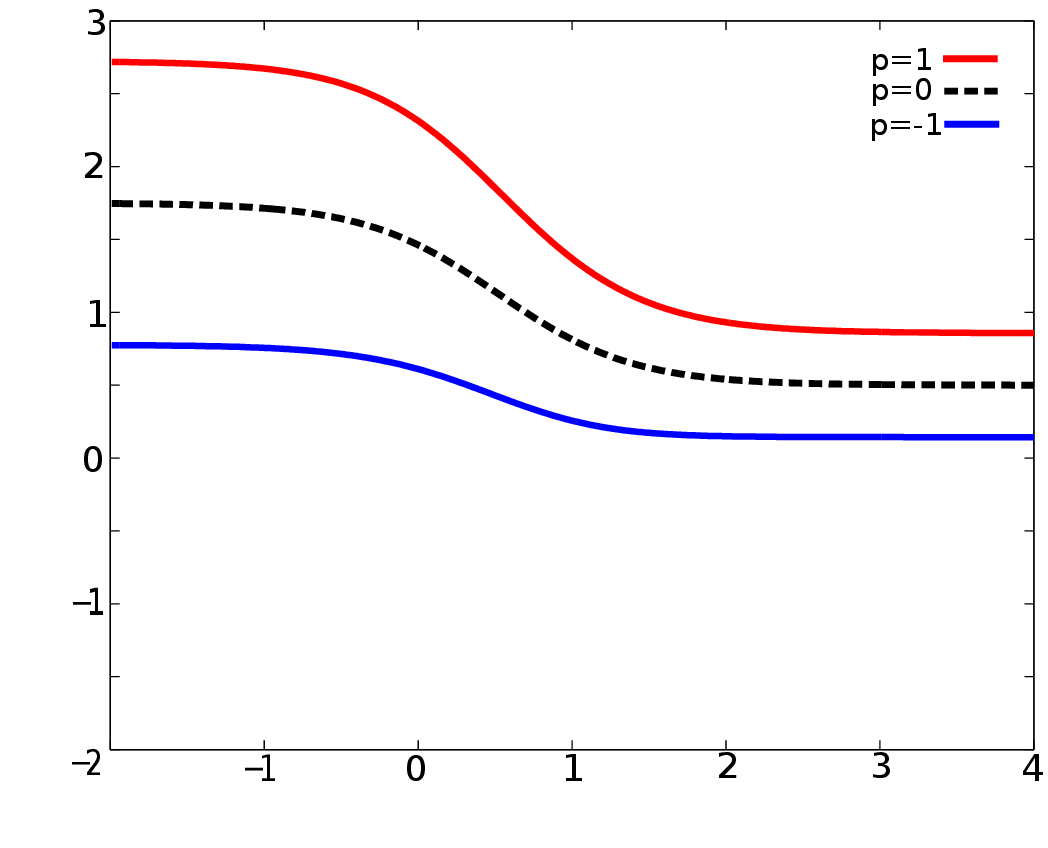}
\caption{The dependence of the branching ratio on logarithm of electron density.
$\gamma_{uv}=0.1\gamma_{vis}$,$\gamma_{uv}=\gamma_{vis}$,$\gamma_{uv}=5\gamma_{vis}$.} \label{br}
\end{figure}

The branching ratio given by Eq. (\ref{Ratio5}) is now the subject of the analysis with respect to the experimental observations in Ref. \cite{YChung}. First
of all we use steady-state values of populations given by solutions (\ref{BBc}) and (\ref{rraa}) in Eq. (\ref{Ratio5}) and obtain \be
R=\gamma_{vis}\left(\frac{4}{r_e}+\frac{1}{r_{vis}}\right) \label{Ratio6}\ee

The common factor $\rho_{cc}$ has been canceled out. According to the experimental observations in Ref. \cite{YChung}, the branching ratio was measured as the
function of electron density. One order of magnitude increase in the electron density resulted in one order of magnitude decrease in the branching ratio. Let
us see how such dependence emerges from the above equation.

The decay constant $\gamma_{vis}$ depends nontrivially on the frequency and the dipole moment, see Eq. (\ref{gam}). Within the range of change of the
experimental conditions, no dependence of the transition frequency (i.e., no line shift) on electron density was found. Some theoretical efforts were made in
Ref. \cite{Chung} to take into account the effect of the screening of the atomic potential by the surrounding plasma. This effect can, in principle, modify the
Coulomb field experienced by the valence electron and, thus, significantly change the transition probability (i.e. dipole moment). However, this change becomes
of an appreciable value only for concentrations much higher than that used in the experiment. So, we conclude that no change takes place in the value of
$\gamma_{vis}$ when concentration varies.

The observed changes of the branching ratio R in Ref. \cite{YChung} can be attributed mainly to the dependence of collisional rates $r_e$ and $r_{vis}$ (and,
of course $r_{UV}$) on the density of free electrons. Apparently the higher the density the more frequently the collision events occur and therefore the larger
the rates. More precisely, we assume linear dependence of the collisional rates on the concentration ($r_i \propto N$) and write for a constant electron
temperature \be r_i(N)=k_i N \label{ri}\ee where $i = e, vis, UV$, and $k_i$ is the proportionality coefficient. Typical range of concentrations where the
effect occurs covers the region $10^{18} - 10^{19} cm^{-3}$. Applying dependences given by Eq. (\ref{ri}) to Eq. (\ref{Ratio6}), we perform simulation and
present obtained results in Fig. \ref{br}. We thus illustrate main result of our study - the sensitive dependence of the branching ratio on the concentration.
The range of the change is of the same order as was observed in the experiments.

For completeness we analyze the dependence of the collisional rates (and therefore, the branching ratio) on the electron temperature of plasma. This
dependence, even when present, cannot be deduced from the experiments in Ref. \cite{YChung} where the electron temperature was estimated as constant under
operating conditions and equal to $\approx 5 eV$. However, in the set of related experiments on measurements of the branching ratio in high-density plasmas of
CIII performed in a later work in Ref. \cite{Glenzer}, the temperature varied considerably and, therefore, we can expect a well pronounced dependence of the
branching ratio on the temperature. Thus for the concentration $0.7 \times 10^{18} cm^{-3}$, the temperature was 5.7 eV; while for higher concentration $2.6
\times 10^{18} cm^{-3}$, the temperature increased to $9.3 eV$.

It is instructive to analyze the simultaneous electron density and temperature dependences. Free electrons in plasmas obey the classical Boltzmann distribution
thus the number of particles $dN$ (in a unit volume) within the range of energies $E + dE$ yields

\be d N=N\frac{2}{\sqrt{\pi}}\frac{1}{(kT)^{3/2}}e^{-\frac{E}{kT}}\sqrt{E}dE\ee

So, for collisional rates we can write

\be r_i(N,T)=2N\overline{k}_i\sqrt{\frac{2kT}{\pi M}}e^{-\frac{E_i}{kT}} \label{rint}\ee
where $\overline{k}_i$ is a cross-section. Then the coefficient $k_i$ is given by

\be k_i=2\overline{k}_i\sqrt{\frac{2kT}{\pi M}}e^{-\frac{E_i}{kT}} \label{ki}\ee

The branching ratio was measured in Ref. \cite{Glenzer} for five concentration/temperature points.

\section{Simulation results}
We have performed numerical simulation to demonstrate the dependency of branch ratio on electron density.  For purposes of plasma diagnostics, we
simulate  the situation with or without the coherent effects due to electron impacts, depending on the relative orientation of the dipole moments of
optical and UV transitions.

Here we assume that $\gamma_{vis}=1$, so that all the other parameters are normalized by $\gamma_{vis}$. The collision-induced incoherent pumping
rates are  dependent on electron density; here we use $r_{UV}=0.001\times N_e$, $r_e=0.1\times N_e$, and $r_v=0.3\times N_e$. We considered three
different UV decay rates with $\gamma_{UV}=0.1,1,5$ to investigate the dependence of branching ratio on electron density.

Fig.\ref{br}.(a) corresponds to small UV decay rate $\gamma_{UV}=0.1\gamma_{vis}$. Fig.\ref{br}.(b,c) correspond to larger UV decay rate $\gamma_{UV}=1,5$
respectively. In each figure, the top red curve is for $p=1$, the bottom blue curve is for $p=-1$, and the dashed line is for $p=0$.

Through these nine curves, we notice that, at large electron density, the dependence of the branching ratio on density for different p factors behave
in a  similar way and trend to $0.5$. In the special case with $p=0$, this process gets to limit much more quickly. It is also clear that with larger
UV decay rate, the dependence of the branching ratio on density for different p factors also behaves very similarly, but at a lower decay rate. At
both sides of the extremely low or high electron density, there clearly limits. Outside this range, the electron density has no effect on the
branching ratio.

Obvious conclusion from the comparison of curves is that the increase of temperature greatly suppresses variations in the branching ratio with concentration.
The relatively small slope of the curve could be the reason or one of the reasons that yielded the very weak concentration dependence and led the experimenters
in Ref. \cite{Glenzer} to the conclusion of the absence of the dependence of the branching ratio on the concentration.

We thus complete the comparison with the experiments. On our way to the main result expressed by formula (\ref{Ratio6}) we made two important assumptions. One
is the condition of maximal coherence. The other is the domination of collisional rates over all relevant decays, see Eq. (\ref{rrgg}). It is instructive now
to see what happens if we relax the first assumption and consider the opposite case - the case of no coherence. Simply set p factor to zero; that is equivalent
to setting $\rho_{ab} = 0$ (at least in the steady-state). Given inequality (\ref{rrgg}), two upper states $|a\ket$ and $|b\ket$ are populated equally in
steady state. In the case of no coherence, the population of the bright state is simply one half of the sum of $\rho_{aa}$ and $\rho_{bb}$. So, we get \be
\rho_{aa}\approx\rho_{BB}\approx(1+\frac{r_e}{2r})\rho_{cc}\ee

With these solutions in Eq. (\ref{Ratio5}), we finally obtain \be R\approx2\frac{\gamma_{vis}}{\gamma_{UV}} \label{Ratio7}\ee for the branching ratio for the
visible and ultraviolet transitions in the case of no coherence between two upper states $|a\ket$ and $|b\ket$. This value is considerably larger than the
branching ratio obtained for the case of maximal coherence, see Eq. (\ref{Ratio6}). Moreover, this value does not depend on collisional rates and therefore on
concentration. The comparison of branching ratios calculated for these two cases demonstrates the key role of coherence in interpreting experimental results
obtained in Ref. \cite{YChung}.

\begin{figure}[t]
\centering
\includegraphics[width=0.4\columnwidth]{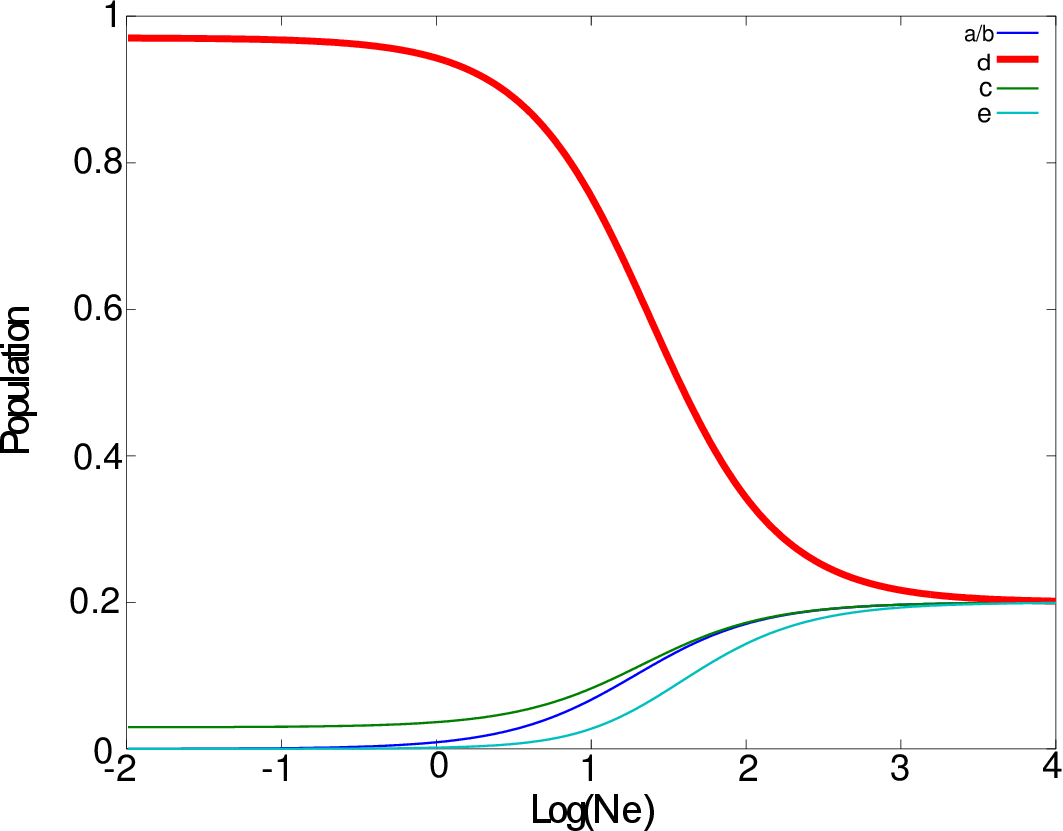}
\includegraphics[width=0.4\columnwidth]{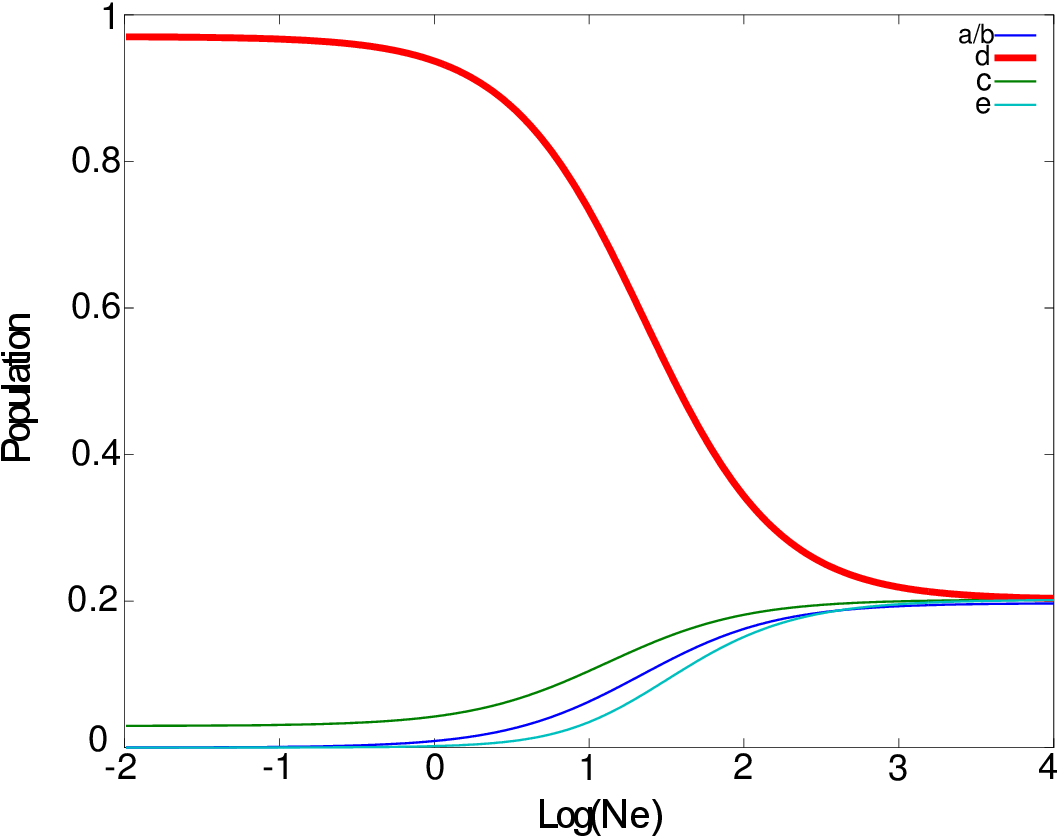}
\includegraphics[width=0.4\columnwidth]{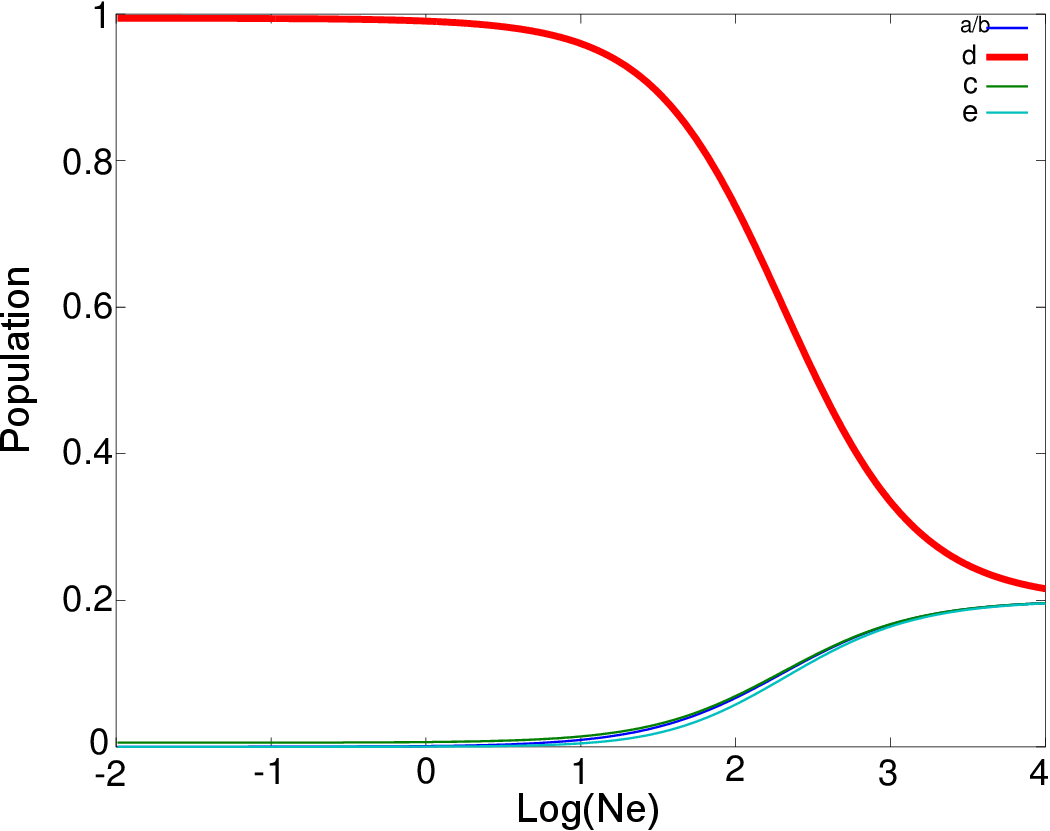}
\includegraphics[width=0.4\columnwidth]{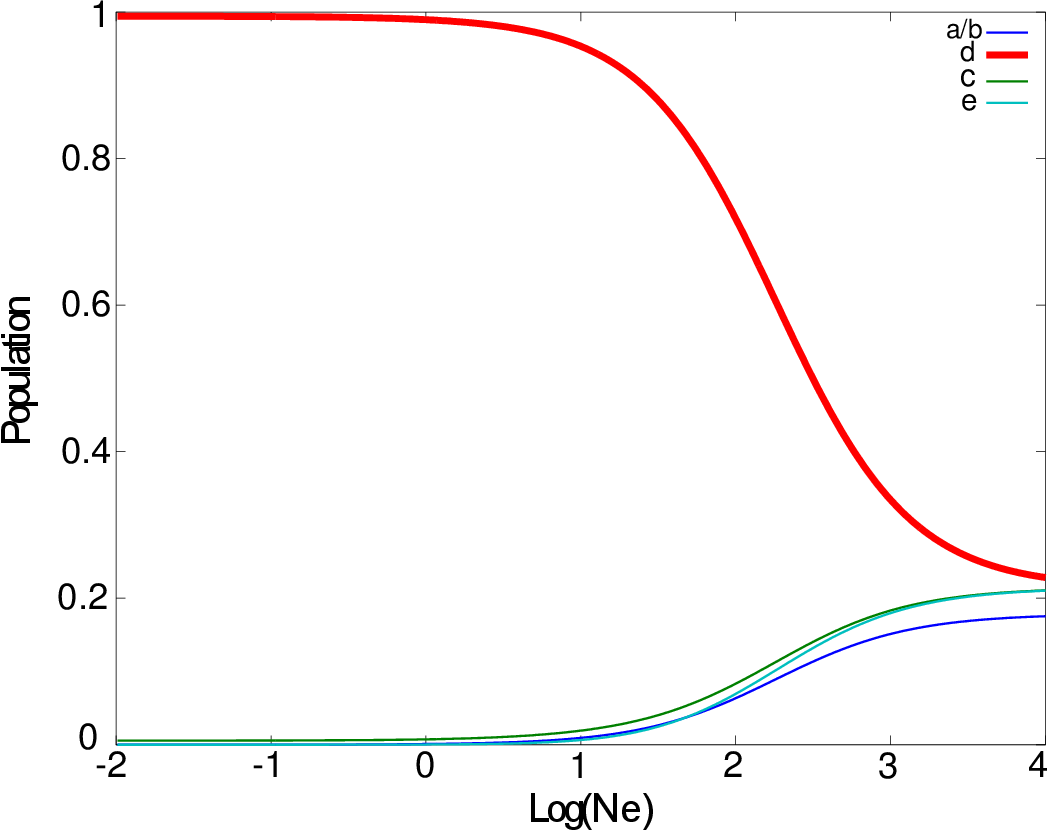}
\includegraphics[width=0.4\columnwidth]{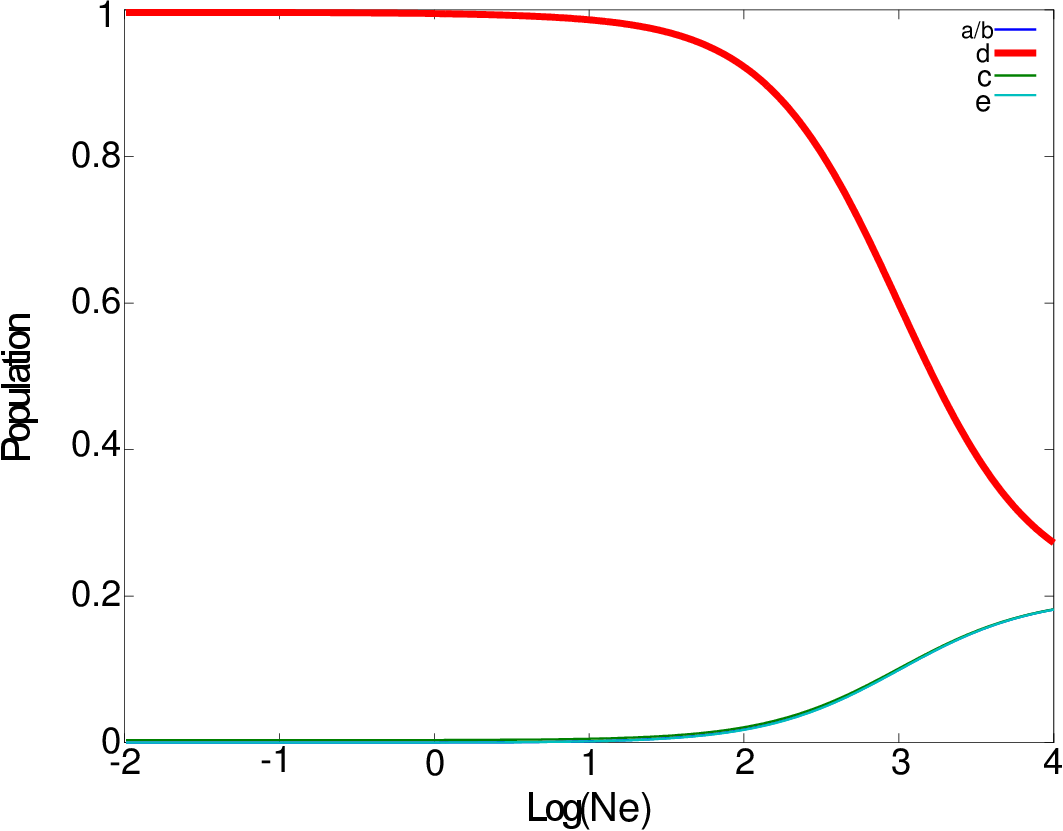}
\includegraphics[width=0.4\columnwidth]{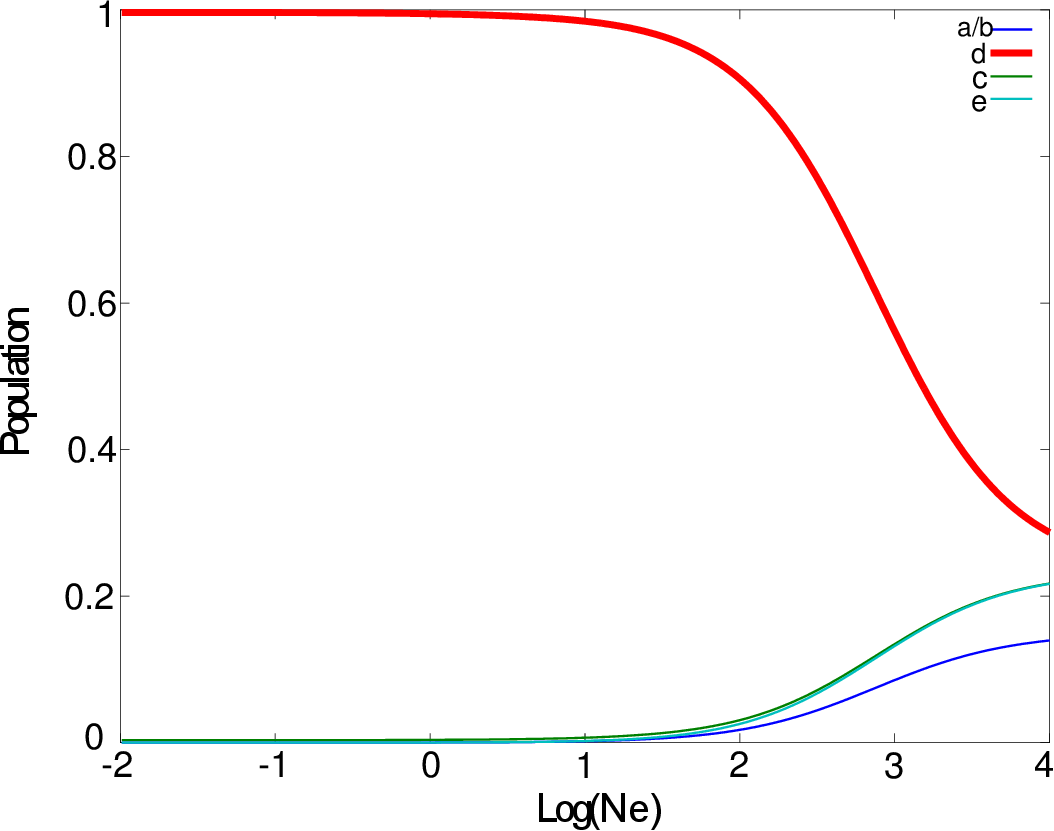}
\caption{The dependence of the populations on logarithm of electron density.$\gamma_{uv}=0.1\gamma_{vis}$, $p=0$, $\gamma_{uv}=\gamma_{vis}$, $p=0$,
$\gamma_{uv}=5\gamma_{vis}$, $p=0$} \label{populations}
\end{figure}
\section{Analytical solutions}
Analytical solutions can be derived from dynamical equations of population densities, ie Eq. (\ref{raa}-\ref{rcb}).We obtain \be
\rho_d=\frac{r_{uv}+\gamma_{uv}}{r_{uv}}\rho_a=R_{da}\rho_a \ee

\be \rho_e=\frac{r_e}{r_e+2\gamma_e}\rho_c=R_{ec}\rho_c \label{e} \ee

\be \rho_a=\rho_b \label{ab}\ee

\be \rho_{ab}=\frac{Pr_v}{\Gamma_{ab}}(\rho_c-\rho_a) \label{L12} \ee where \be \Gamma_{ab}=r_v+g_v+g_{uv}/2 \ee

and \be \rho_c=R_{ca}\rho_a \ee where we introduce \be
R_{ca}=\frac{2(r_v+g_v-P^2r_v^2/\Gamma_{ab})}{2r_v+r_e-r_eR_{ec}-2P^2r_v^2/\Gamma_{ab}} \ee

Here we can use the conservation of the electron number. Suppose that the total number of electrons in all levels is \be \rho_a+\rho_b+\rho_c+\rho_d+\rho_e=1
\label{number} \ee

Thus we get \be \rho_a=\frac{1}{2+R_{ca}+R_{da}+R_{ec}R_{ca}} \ee

By exploiting the relationships of density between different levels, we can find each level's density.

It is known that the density of the bright state is defined as

\begin{equation}
\rho_{BB}=\rho_a+\rho_b+2\rho_{ab}
\end{equation}
and the definition of the branch ratio is
\begin{equation}
R\approx\frac{\rho_{BB}}{\rho_a}
\end{equation}

By using all the results we have obtained, we finally get the expression of the branch ratio

\begin{equation}
R=2+\frac{2Pr_v}{\Gamma_{ab}}(2R_{ca}-1)
\end{equation}
It can be noticed that the branching ratio Limit, at low and high electron densities is

\be R=\left\{\begin{array}{cc}
                        1+2\frac{\gamma_v}{\gamma_{UV}} & N_e\rightarrow 0 \\
                        2\frac{\gamma_v}{\gamma_{UV}} & N_e\rightarrow \infty
                      \end{array} \right.\ee

With no coherence excited by electron collisions, the ratio is given by \be R=\frac{\gamma_v(\gamma_{UV} + 2(\gamma_v + r_v))}{\gamma_{UV}(\gamma_{v} +
r_v)}\ee

\section{Discussion}

So far we discussed the suppression of spontaneous decay on the visible transition. This suppression is due to the two-photon coherence
induced by the interference of collision processes along two overlapping optical transitions $|a\ket-|c\ket$ and $|b\ket-|c\ket$. In order
to quantify the degree of the suppression we introduced the branching ratio as the ratio of total spectral-line intensities for the two
transitions. Here, one transition is the visible transition (actually, simultaneously two transitions) of interest and the other transition
is the reference ultraviolet transition. This branching ratio calculated for the case of maximal two-photon coherence, see Eq.
(\ref{Ratio6}), is compared to the branching ratio evaluated for the case of no coherence, see Eq. (\ref{Ratio7}). The degree of the
coherence-induced suppression can be deduced as the ratio of these two branching ratios.

This study demonstrates the possibility of coherent effects in plasmas, where the coherence is induced by the interference of incoherent
processes. Here the collisions of free electrons with ions represent these incoherent processes, and they indeed happen rather frequently
for concentrations as high as $10^{18} cm^{-3}$. When they dominate over relevant decay rates, the two-photon coherence becomes of
substantial value that leads to efficient suppression of the spontaneous emission on the visible transition. This suppression was
registered experimentally and reported by Chung, Lemaire, and Suckewer in Ref. \cite{YChung}. Moreover, we are able to explain sensitive
dependence of the degree of suppression on concentration of free electrons. It is interesting that the dependence on the concentration only
appears due to the collision-induced coherence and disappears in the absence of the coherence, as comparing Eqs. (\ref{Ratio6}) and
(\ref{Ratio7}) shows.

\section{Acknowledgement}
We thank Marlan Scully for his support on this work. This work is supported by the Robert. A. Welch Foundation (Grant A-1261), D. S. thanks the
support from the Fujian Provincial Natural Science Foundation(2018I0019), Science Technology innovation project of Xiamen(3502Z20183062),Y.R.
gratefully acknowledges the support from the UNT Research Initiation Grant.

\end{document}